\documentstyle[12pt,epsfig]{article}
\let\section=\subsection     \let\subsection=\subsubsection                
  
\begin{document}
\begin{center}
  {\large \bf INFLUENCE OF DIFFERENT COLLECTIVE}\\[2mm]
  {\large \bf COORDINATES}\\[2mm]
  {\large \bf ON SPONTANEOUS FISSION PROCESS}\\[2mm]
  {\large \bf IN Fm ISOTOPES\footnote{ Contribution to the International
  Workshop XXIV on Gross Properties of Nuclei and Nuclear Excitations,
  Hirschegg, Austria, Jan.15--20, 1996.}
  }\\[5mm]
  A.~STASZCZAK and Z.~\L{}OJEWSKI \\[5mm]
  {\small \it  Department of Theoretical Physics, M. Curie--Sk\l{}odowska
  University \\
  pl. M. Curie--Sk\l{}odowskiej 1, 20--031 Lublin, Poland\\[8mm] }
\end{center}

\begin{abstract}\noindent
We study the role of different collective degrees of freedom in
spontaneous fission process of even--even Fm isotopes. To find a proper
collective space we examine a reach collection of nuclear shape
parameters $\{\beta_{\lambda}\}$, with $\lambda$=2,3,4,5,6 and 8; as
well as the paring degrees of freedom i.e. proton $\Delta_{p}$ and
neutron $\Delta_{n}$ pairing gaps. On the basis of the multidimensional 
dynamic--programming method (MDPM) the optimal collective space
$\{\beta_{2}, \beta_{4}, \beta_{6}, \Delta_{p}, \Delta_{n}\}$ 
for Fm isotopes was found. 
\end{abstract}

\section{Introduction}
It is commonly known that experimental values of the spontaneous fission
half--lives $(T_{sf})$ of nine even--even Fm isotopes (N = 142,
144, ..., 158) form approximately two sides of an acute--angled
triangle with a vertex in N = 152. The rapid changes in
$T_{sf}$ on both sides of $^{252}$Fm are particularly dramatic
for the heavier Fm isotopes, where $T_{sf}$ descends by about ten
orders of magnitude when one passes from $^{254}$Fm to $^{258}$Fm.
This strong nonlinear behaviour of $T_{sf}$ {\em vs.} neutron number N 
produces good opportunity for testing theoretical models.

The aim of this paper is to present $T_{sf}$ of even--even Fm isotopes, 
obtained on the basis of a~wholly dynamical analysis in different
multidimensional collective spaces.

The used method is described in sect.~2, the results and discussion
are given in sect.~3 and the conclusions are presented in sect.~4.

\section{Formulation of the Method}

\subsection{Spontaneous Fission Half--Life} 
The spontaneous fission half--lives $T_{sf}$ for the even-even Fm
isotopes were evaluated within the one--dimensional WKB semiclassical 
approximation
\begin{equation}
 T_{sf} [yr] = 10^{-28.4} [1+exp S(L)]\,,
\end{equation}
where $S(L)$ is the action--integral along a fission path $L(s)$ in the
multi--dimensional deformation space $\{X_\lambda\}$
 \begin{equation}
 S(L) = \int^{s_2}_{s_1} \left\{{2 \over \hbar^2} \, B_{\rm eff}(s)
 [V(s) - E]\right\}^{1/2} ds\,.
\end{equation}
An effective inertia associated with the fission motion
along the path $L(s)$ is
\begin{equation}
B_{\rm eff}(s) = \sum_{\lambda,\mu} \, B_{X_\lambda X_\mu} \,
  {dX_{\lambda} \over ds} {dX_\mu \over ds}\,.
\end{equation}

In above equations $ds$ defines the element of the path length in
the $\{X_\lambda\}$ space. The integration limits $s_1$ and
$s_2$ correspond to the entrance and exit points of the barrier
$V(s)$, determined by a~condition $V(s) = E$, where 
$E = V(X^0_\lambda)$ + 0.5~MeV is defined as a sum of a~ground--state
energy $V(X^0_\lambda)$ and a~zero--point energy in the
fission direction at the equilibrium deformation $X^0_\lambda$
and denotes the energy of the fissioning nucleus.

\subsection{Potential Energy and Inertia Tensor}
The potential energy $V$ is calculated by the
macroscopic--microscopic model very similar to 
that used in \cite{SS}. For the
macroscopic part we used the Yukawa--plus--exponential
finite--range model \cite{MN} and for microscopic part the Strutinsky
shell correction, based on the Woods--Saxon single--particle
potential with ``universal'' variant of the parameters \cite{CD}. The
single--particle potential is extended to involve residual
pairing interaction, which is treated in the BCS approximation.
The inertia tensor $B_{X_\lambda X_\lambda}$, which
describes the inertia of the nucleus with respect to change of
its shape, is calculated in the cranking
approximation (cf. e.g. \cite{BP}).

\subsection{Multidimensional Dynamic--Programming Method (MDPM)} 
Dynamic calculations of the spontaneous--fission half--lives
$T_{sf}$ are understood as a quest for least--action
trajectory $L_{\rm min}$ which fulfills a principle of the least--action
$\delta[S(L)]~=~0$. To minimize the action integral (2) we used the
dynamic--programming method \cite{BP}. Originally this method was
used only for two--dimensional deformation space. We extended the
method up to four dimensions.

In opposite to approximation used in \cite{SS,SK}, where only two of
coordinates ($\beta_2$ and $\beta_4$) have been handled dynamically 
and the remaining degrees of freedom have been found only by
minimization of the potential energy V, in our multidimensional 
dynamic--programming method (MDPM) all coordinates are treated
dynamically as independent variables.

\section{Results and Discussion}

\subsection{Effect of Higher Even--Multipolarity $\beta_6$ and
$\beta_8$ Parameters on the Spontaneous Fission Half--Lives} 

\begin{center}
   \hspace{0cm}
   \psfig{file=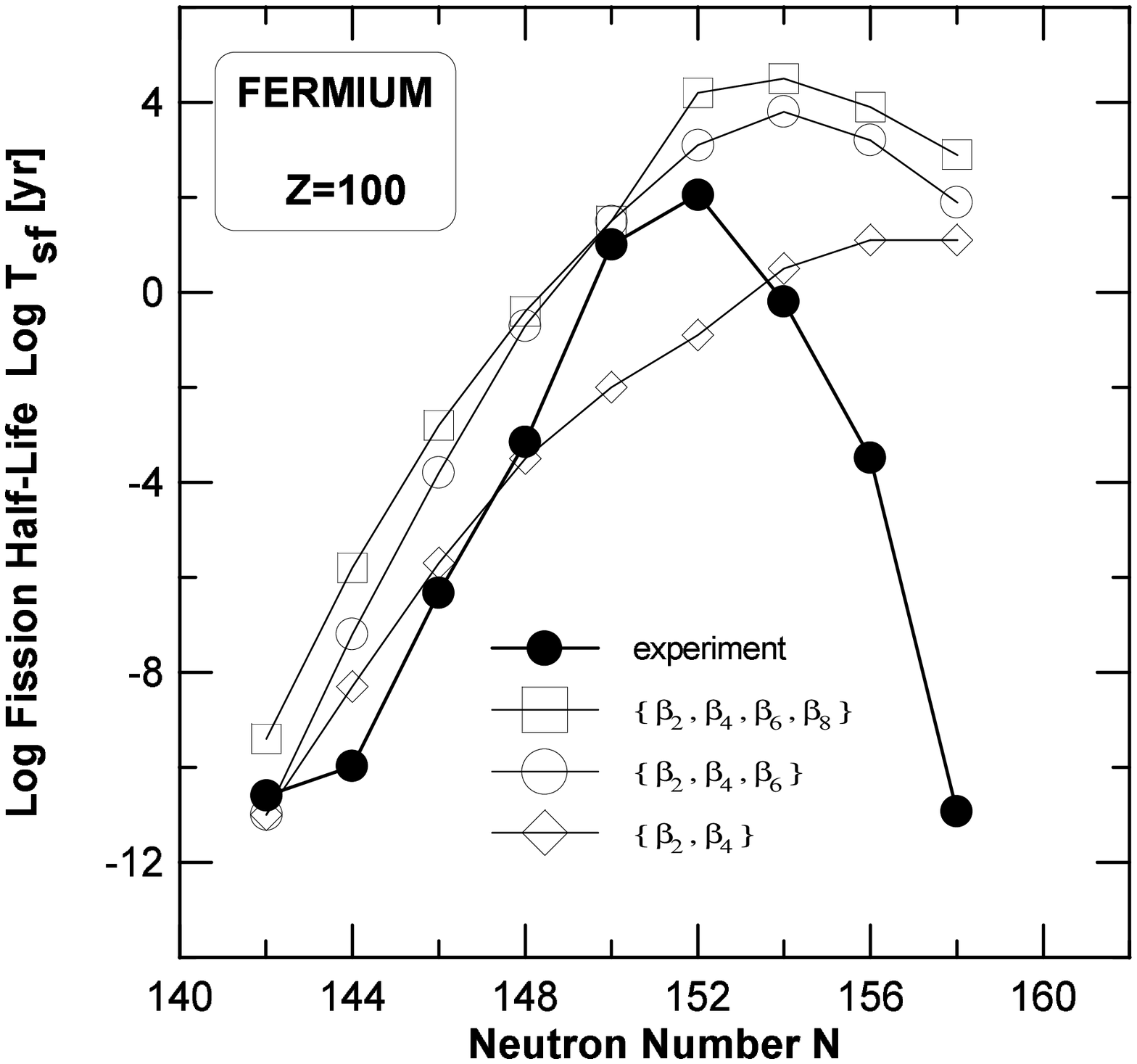,height=11cm}
\begin{minipage}{13cm}
   \vspace{-5.5cm}
\baselineskip=12pt
{\begin{small}
Fig.~1. Logarithms of the calculated in the different collective spaces
spontaneous fission half--lives of the even--even Fm isotopes
{\em vs.} neutron number. The experimental values are shown for comparison.
\end{small}}
\end{minipage}
\end{center}
\vspace{-2.5cm}
To find the proper collective space for description of the fission
process we examine three kinds of deformation spaces:
$\{\beta_{2}, \beta_{4}, \beta_{6}, \beta_{8}\}$, 
$\{\beta_{2}, \beta_{4}, \beta_{35}, \beta_{6}\}$ and 
$\{\beta_{2}, \beta_{4}, \Delta_{p}, \Delta_{n}\}$, where parameter 
$\beta_{35}$ defines an average
trajectory in a ($\beta_3, \beta_5$) plane; $\Delta_{p}$ and $\Delta_{n}$
denotes proton and neutron pairing gap, respectively.

Fig.~1 shows the logarithm of the spontaneous fission half--lives 
$T_{sf}$, given in years, for even--even Fm isotopes;
obtained when only deformations of the even--multipolarities
$\{\beta_{\lambda}\}$, $\lambda$=2,4,6,8 are considered.
One can see that effect of deformation $\beta_6$ on $T_{sf}$ is
stronger then deformation $\beta_8$. This is in agreement with earlier
results (see e.g. \cite{SS}).

\subsection{Role of the Reflection--Asymmetry Shape Parameter 
            $\beta_{35}$}
To examine the role of the deformations with odd multipolarities on 
$T_{sf}$ we collect the $\beta_3$ and $\beta_5$ parameters in one
parameter $\beta_{35}{\equiv}(\beta_3,\beta_5{=}0.8\beta_3)$ and
perform the dynamical calculation of $T_{sf}$ in four--dimensional
collective space $\{\beta_{2}, \beta_{4}, \beta_{35}, \beta_{6}\}$.
Results of this study are presented in Fig. 2. It is easy to see that
reflection--asymmetry shape parameter $\beta_{35}$ does not change 
$T_{sf}$ in Fm region. 
 
The reason of this lies in the dynamical treatment of fission process.
The parameters $\beta_3$ and $\beta_5$ shorten the static fission barriers
(i.e. barriers along static paths), however the effective inertia 
$B_{eff}$, eq. (3), along these static paths are larger than along the path
with $\beta_{35}{=}0$.
\vspace{-0.5cm}
\begin{center}
   \hspace{0cm}
   \psfig{file=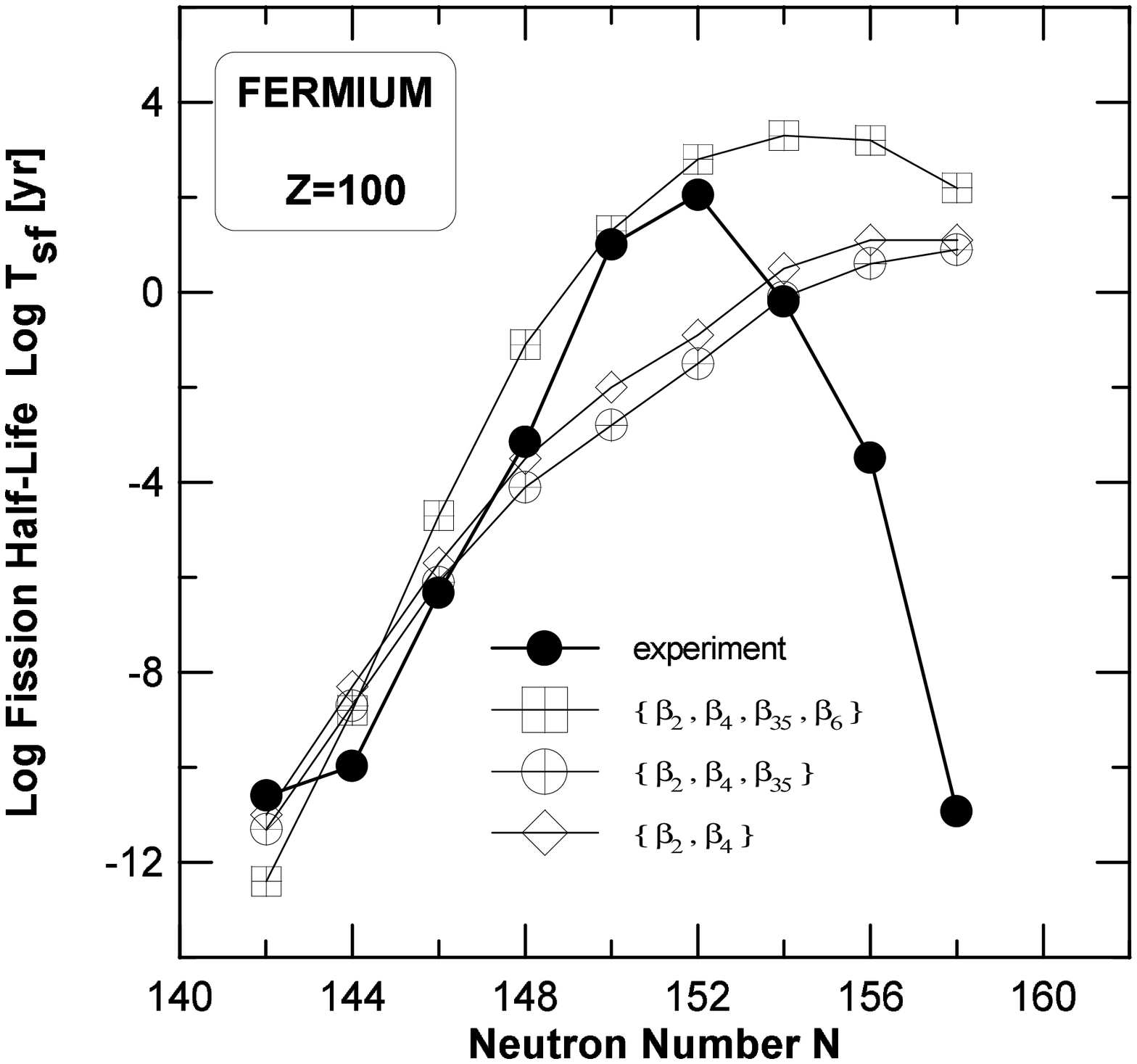,height=11cm}
\begin{minipage}{13cm}
   \vspace{-5.5cm}
\baselineskip=12pt
{\begin{small}
Fig.~2. The same as in Fig.1, but obtained when reflection--asymmetry 
$\beta_{35}$ shape parameter (see text) is included in the collective space. 
 \end{small}}
\end{minipage}
\end{center}
\vspace{-2.5cm}
\vspace{-0.5cm}
\subsection{Influence of the Pairing Degrees of Freedom}
The residual pairing interactions are usually treated in the stationary
way (i.e.in BCS approximation). The idea of dynamical calculations of
$T_{sf}$ in multidimensional collective space spanned by the shape
parameters as well as the pairing--field parameters ($\Delta_p,\Delta_n$)
was proposed in ref. \cite{MB} and practically applied to 
the macroscopic--microscopic calculations with Nilsson single--particle
potential in \cite{SB,SP}
\begin{center}
   \hspace{0cm}
   \psfig{file=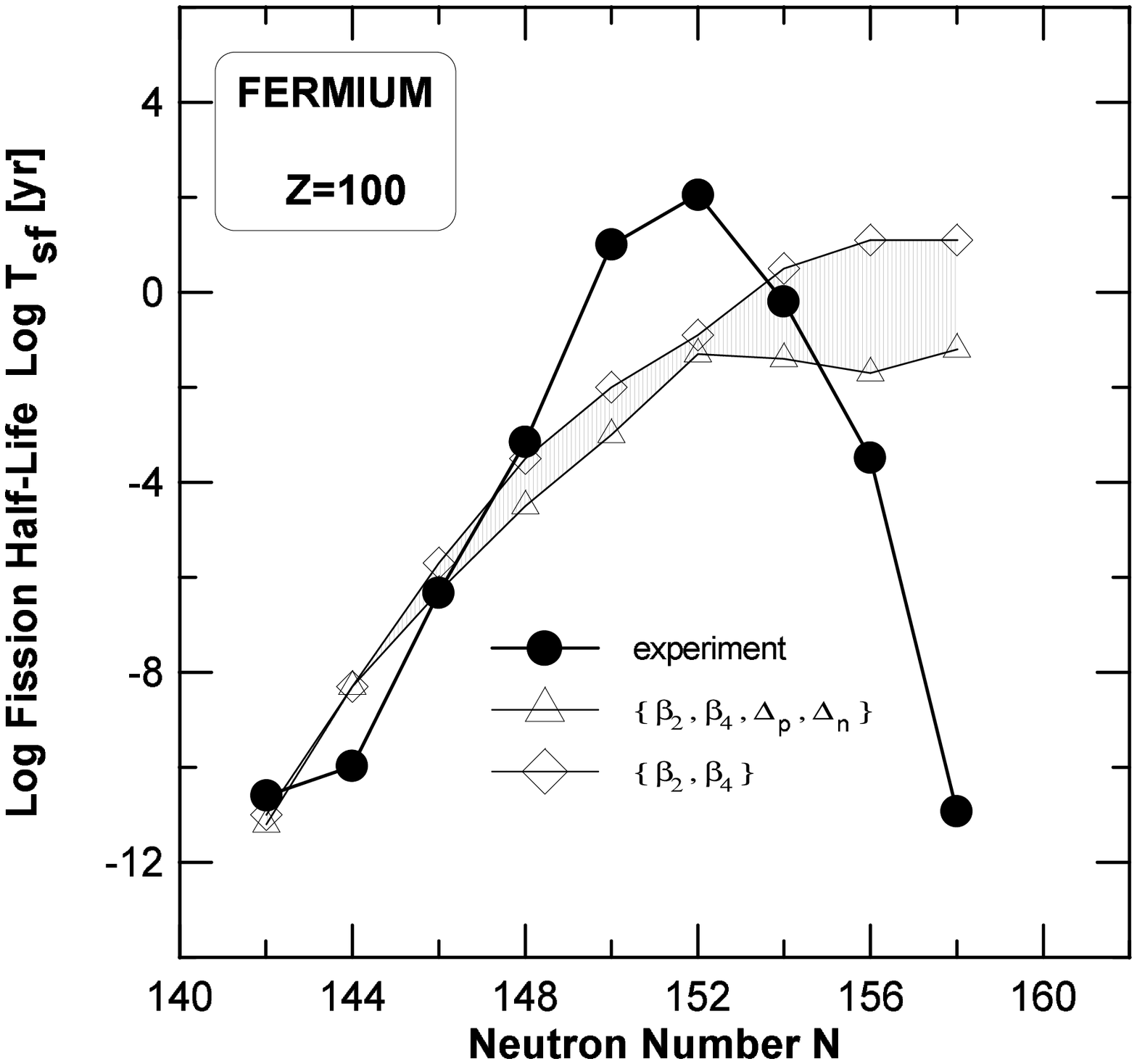,height=11cm}
\begin{minipage}{13cm}
   \vspace{-5.5cm}
\baselineskip=12pt
{\begin{small}
Fig.~3. Comparison between $T_{sf}$ of even-even Fm isotopes, obtained 
in collective spaces with and without pairing degrees of freedom
i.e. proton $(\Delta_p)$ and neutron $(\Delta_n)$ pairing gaps. 
The difference between two kinds of $T_{sf}$ is shadowed.  
 \end{small}}
\end{minipage}
\end{center}
\vspace{-2.0cm}
In Fig. 3 we present $T_{sf}$ of even--even Fm isotopes obtained in
four--dimensional collective space
$\{\beta_{2}, \beta_{4}, \Delta_{p}, \Delta_{n}\}$ and in two--dimensional
space $\{\beta_{2}, \beta_{4}\}$. The difference between $T_{sf}$
got with and without pairing degrees of freedom is shadowed in the figure.

This difference represents the dynamical effect of the pairing degrees
of freedom on the spontaneous fission half--lives $T_{sf}$.

As it is seen, proton $(\Delta_p)$ and neutron $(\Delta_n)$ pairing gaps 
reduce $T_{sf}$ for Fm isotopes with N $>$ 152 for about 3 orders of
magnitude.
\vspace{0.0cm}
\subsection{Optimal Multidimensional Deformation Space}
Fig.~4 presents $T_{sf}$ of even-even Fm isotopes calculated in
$\{\beta_2, \beta_4, \beta_6\}$ collective space and corrected by the 
effect of the pairing degrees of freedom from Fig.~3.

This correction is strongly isotopic dependent and considerably
improves theoretical prediction of $T_{sf}$. 
The dynamical study in different deformation spaces of various dimensions
has shown that optimal collective space for description of
spontaneous fission half--lives in even--even Fm isotopes 
is five-dimensional space
$\{\beta_{2}, \beta_{4}, \beta_{6}, \Delta_{p}, \Delta_{n}\}$. 
\vspace{-0.5cm}
\begin{center}
   \hspace{0cm}
   \psfig{file=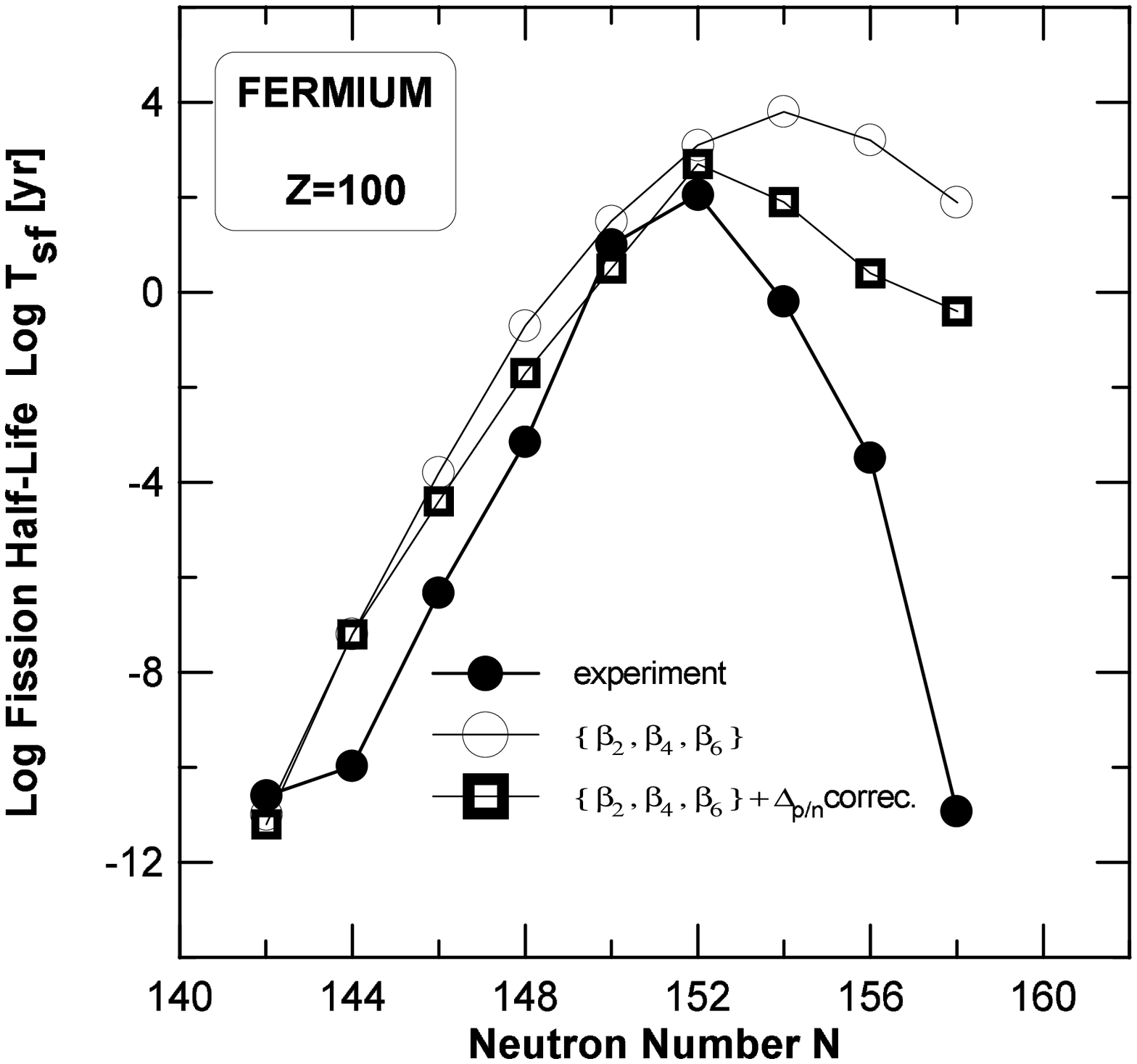,height=11cm}
\begin{minipage}{13cm}
   \vspace{-5.5cm}
\baselineskip=12pt
{\begin{small}
Fig.~4. Logarithms of $T_{sf}$ of even-even Fm isotopes from Fig.~1 
calculated in $\{\beta_2, \beta_4, \beta_6\}$ collective space and
corrected by the effect of the pairing degrees of freedom from Fig.~3.
 \end{small}}
\end{minipage}
\end{center}
\vspace{-2.5cm}
\section{Conclusions}
The following conclusions may be drawn from the present study.
\begin{enumerate}
\item The contribution of parameter $\beta_8$ to $T_{sf}$ is negligible.
\item In dynamical calculations the deformations with odd
      multipolarities $\beta_3$ and $\beta_5$ do not change $T_{sf}$.
\item The proton $\Delta_p$ and neutron $\Delta_n$ pairing gaps 
      reduce $T_{sf}$ for heavy even--even Fm isotopes.
\item The optimal collective space for description of spontaneous fission
      half--lives in even--even Fm isotopes is 
      $\{\beta_{2}, \beta_{4}, \beta_{6}, \Delta_{p}, \Delta_{n}\}$.
\end{enumerate}

\end{document}